\documentclass[11pt,reqno]{article}
\usepackage[english]{babel}
\usepackage{amsmath,amsthm}
\usepackage{amsfonts}
\usepackage{graphicx}
\usepackage{enumerate}
\usepackage[usenames,dvipsnames]{color}
\usepackage{subfigure}
\usepackage[right,pagewise,displaymath, mathlines]{lineno}
\usepackage{epstopdf}
\usepackage{color}
\usepackage{multirow}
\usepackage[colorlinks = true,urlcolor = blue,citecolor = blue,breaklinks]{hyperref}

\usepackage{natbib}
\usepackage{dsfont}
\usepackage{amssymb}

\usepackage{pgfplots}
\usepackage{amssymb}
\usepackage{url}
\usepackage{fancyhdr}
\usepackage{indentfirst}
\usepackage{titlesec}
\usepackage{dsfont}
\usepackage[misc]{ifsym}
\usepackage{booktabs}

\def\P{\mathbb{P}}

\def\E{\mathbb{E}}

\def\R{\mathbb{R}}

\def\X{\mathcal{X}}

\def\I{\mathcal{I}}

\def\id{\mathds{1}}

\def\X{\mathcal{X}}

\def\d{\,\mathrm{d}}

\newcommand{\VaR}{\mathrm{VaR}}
\newcommand{\ES}{\mathrm{ES}}

\DeclareMathOperator*{\argmin}{arg\,min}

\usepackage{geometry}
\usepackage[onehalfspacing]{setspace}

\numberwithin{equation}{section}
\numberwithin{figure}{section}
\numberwithin{table}{section}

\newtheorem{theorem}{Theorem}[section]

\newtheorem{lemma}{Lemma}[section]

\newtheorem{proposition}{Proposition}[section]

\theoremstyle{definition}
\newtheorem{definition}{Definition}[section]
\newtheorem{example}{Example}[section]

\theoremstyle{remark}

\numberwithin{equation}{section}

\begin{document}

%
%
%
%
%
%
%
%
%
%
%
%
%
%
%
%
%

\title{Risk measures induced by efficient insurance contracts}

\author{Qiuqi Wang\thanks{Department of Statistics and Actuarial Science, University of Waterloo,  Canada. \Letter~{\scriptsize \url{q428wang@uwaterloo.ca}}}  \and	Ruodu Wang\thanks{Department of Statistics and Actuarial Science, University of Waterloo,  Canada. \Letter~{\scriptsize\url{wang@uwaterloo.ca}}} \and Ri\v cardas Zitikis\thanks{School of Mathematical and Statistical Sciences, University of Western Ontario,  Canada. \Letter~{\scriptsize\url{rzitikis@uwo.ca}}}}

\maketitle

\begin{abstract}
The Expected Shortfall (ES) is one of the most important regulatory risk measures in finance, insurance, and statistics, which has recently been characterized via sets of axioms from perspectives of portfolio risk management and statistics. Meanwhile, there is large literature on insurance design with ES as an objective or a constraint. A visible gap is to justify the special role of ES in insurance and actuarial science. To fill this gap, we study characterization of risk measures induced by efficient insurance contracts, i.e., those that are Pareto optimal for the insured and the insurer. One of our major results is that we characterize a mixture of the mean and ES as the risk measure of the insured and the insurer, when contracts with deductibles are efficient. Characterization results of other risk measures, including the mean and distortion risk measures, are also presented by linking them to different sets of contracts.
	
~

	\end{abstract}

\section{Introduction}

Optimal insurance and reinsurance design problems have been a prevalent topic for both researchers and practitioners in insurance for decades, since the seminal work of   \cite{A63} showing that deductible insurance is optimal for a risk-averse insured when the insurer is risk neutral.

Recent studies  on optimal (re)insurance design problems have shown considerations  from several different perspectives. The majority of the studies focus on optimization under specific classes of optimization criteria quantifying the risk of decision makers; see e.g., \cite{S81} for expected utilities; \cite{GS96} and \cite{S97} for criteria preserving second-order stochastic dominance; \cite{CT07}, \cite{CTWZ08} and \cite{BT09} for Value-at-Risk (VaR) and  the Expected Shortfall (ES, also called CTE or TVaR in the above literature);
\cite{CYW13} for distortion risk measures or dual utilities \citep{Y87}; and \cite{BM04} for regret-theoretical expected utilities.
For more recent developments on optimal insurance with risk measures, we refer to \cite{CC20} and the references therein.
Moreover, optimal (re)insurance contract design problems are studied under a variety  of constraints and formulations. \cite{G96} derived an optimal form of insurance contracts when wealth loss is divided into insurable and uninsurable components. \cite{CM04} examined the demand for insurance with indemnities constrained by upper limits. More recently,
\cite{HHN19} studied optimal insurance policy indemnity schedules with limited liability and background risk.
\cite{LTT21} analyzed the set of universally marketable indemnities with risk measures preserving convex orders.

Most of the previous literature aims to derive optimal forms of ceded loss functions under various scenarios and constraints. To the best of our knowledge, there is  no relevant research on (re)insurance contract design problems focusing on  identifying risk measures adopted by the insured and the insurer.  Therefore, we study optimal insurance contract design problems through a distinctive perspective if compared to previous literature. Namely, the main goal of the present paper is to answer the following (converse) question: In order for efficient contracts to be some sets of contracts commonly seen in insurance practice (e.g., of deductible form), which risk measures should the insurer and the insured use? Specifically, we characterize different classes of risk measures adopted by the insured and the insurer given different sets of ceded loss functions that are Pareto optimal.

The risk measure ES has been widely applied in the contexts of financial regulation, risk management, and insurance. In particular, there is a growing academic literature on various problems using ES   in actuarial science (where ES is often called TVaR).
Most of these studies motivate the use of ES as a coherent risk measure \citep{ADEH99} and its advantages over the risk measure Value-at-Risk. Recently, \cite{WZ21} proposed the axiom called ``no reward for concentration" (NRC) which, together with a few other standard axioms, characterizes ES. The main objective of \cite{WZ21} is to separate ES from other coherent risk measures via the axiom of NRC, thus answering the question of why one uses ES instead of other risk measures from an axiomatic point of view. The interpretation and implication of the NRC axiom in financial regulation have been extensively discussed in \cite{WZ21} and they are well motivated from the perspective of the Fundamental Review of the Trading Book by \cite{BCBS2016, BCBS2019}; see also an alternative formulation for axiomatizing ES in \cite{HWWW21}.

Given the big volume of research with ES in actuarial science, it is of great interest to understand whether ES plays a special role in insurance.
The NRC axiom of \cite{WZ21} does not apply in the insurance context since it is interpreted as a requirement of portfolio risk assessment. To understand the special role of ES in insurance,  new insights that are specific to insurance design are therefore needed.  

We work mainly within the framework of convex risk measures of \cite{FS02}, which is a flexible and popular class of risk measures in risk management. 
As the main contribution of this paper, we show that
the set of efficient ceded loss functions of deductible form corresponds to the family of mixtures of ES and  the mean (Theorem \ref{thm:ES}).
If we further impose lower semicontinuity as in \cite{WZ21}, then we arrive at the family of ES (Lemma \ref{lem:mES}). 
Our work also extends  \cite{EMWW21}, who characterized the mixture of the mean and ES,   called an $\ES/\E$-mixture, as the only coherent Bayes risk measure from the perspective of statistical inference.
In addition,   if the set of efficient ceded loss functions is the set of all slowly growing  ($1$-Lipschitz) functions, then the corresponding risk measures are precisely the convex  distortion risk measures (Theorem \ref{th:1}). Mathematically, our results are based on connecting various risk measures with different additivity forms over the ceded losses and the retained losses.

For  illustrative purposes, we take the perspective of an insurance design problem between an insurer and an insured. Our technical results  can certainly be applied in the reinsurance setting as well, where risk measures are often encountered.  

The rest of the paper is organized as follows. Section \ref{sec:risk} contains some preliminaries on insurance losses and risk measures. Section \ref{sec:design} sets up the formulation of the insurance contract design problem and states economic assumptions. Section \ref{sec:char} contains our main characterization results of the risk measures used by the insured and the insurer given different Pareto-optimal sets of ceded loss functions. The results make natural connections between some common sets of ceded loss functions   and common classes of risk measures  in insurance practice. We also discuss economic implications of these results on the design of insurance menus by the insurer. Section \ref{sec:tech} contains proofs of the main results accompanied with relevant technical lemmas.

\section{Preliminaries on risk measures}\label{sec:risk}

We consider a probability space $(\Omega,\mathcal{F},\P)$. Let $\X$ be the set of all bounded random variables, and let $\X_+$ be the set of all non-negative random variables in $\X$ representing insurable losses.
Let $\mathcal I$ be a class of non-negative functions on $[0,\infty)$ which represent possible insurance ceded loss functions. 
For an insurable loss random variable $X \in \X_+$ and a contract $f\in \mathcal I$,
$f(X)$ represents the payment to the insured, and $X-f(X)$ represents the retained loss of the insured. 
Losses are usually quantified by risk measures which are mappings from $\X$ to the set of real numbers, representing riskiness. Below we recall some properties of risk measures $\rho$, which are commonly encountered in the risk management literature.
\begin{itemize}
\item[] \emph{Law invariance}: $\rho(X)=\rho(Y)$ for all $X,Y\in\X$ such that $X\stackrel{\mathrm{d}}{=}Y$.\footnote{We write $X\stackrel{\mathrm{d}}{=}Y$ when two random variables $X$ and $Y$ follow the same distribution.}
\item[] \emph{Monotonicity}: $\rho(X)\ge\rho(Y)$ for all $X,Y\in\X$ such that $X\ge Y$.
\item[] \emph{Translation invariance}: $\rho(X+d)=\rho(X)+d$ for all $X\in\X$ and $d\in\R$.
\item[] \emph{Convexity}: $\rho(\lambda X+(1-\lambda)Y)\le\lambda\rho(X)+(1-\lambda)\rho(Y)$ for all $X,Y\in\X$ and $\lambda\in[0,1]$.
\item[] \emph{Positive homogeneity}: $\rho(\lambda X)=\lambda\rho(X)$ for all $X\in\X$ and $\lambda\ge 0$.
\end{itemize}
Following \cite{ADEH99} and \cite{FS16}, $\rho$ is a \emph{monetary risk measure} if it is monotone and translation invariant; a monetary risk measure $\rho$ is called a \emph{convex risk measure} if it satisfies convexity, and it is \emph{coherent} if it is also positively homogeneous.\footnote{\cite{ADEH99} defined coherent risk measures     via subadditivity instead of convexity. A risk measure  $\rho$ is subadditive if $\rho(X+Y)\le \rho(X)+\rho(Y)$ for all $X,Y\in\X$. Subadditivity and convexity are equivalent when positive homogeneity holds.}
For $X\in\X$, a \emph{distortion risk measure} is defined as
$$\rho(X)=\int^\infty_0 h(\P(X > x ))\d x + \int_{-\infty}^0 (h (\P (X> x))-1)\d x,$$
where $h:[0,1]\to[0,1]$ is an increasing function with $h(0)=0$ and $h(1)=1$, and $h$ is called the \emph{distortion function} of $\rho$. Distortion risk measures are always monetary, positively homogeneous, and law invariant, and they are coherent if and only if their distortion functions are concave; see e.g., \cite{WWW20b}. For the application of distortion risk measures to insurance premium principle calculation, see \cite{WYP97}.
For   $X\in\X$ and $p\in(0,1)$, the \emph{Value-at-Risk}  (VaR) is the left-quantile given by
$$\VaR_p(X)=F^{-1}_X(p)=\inf\{x\in\R:\P(X\le x)\ge p\}.$$
For   $X\in\X$ and $p\in[0,1)$, the \emph{Expected Shortfall} (ES) is defined as 
$$\ES_p(X)=\frac{1}{1-p}\int^1_p\VaR_t(X)\d t.$$
It is well known that $\ES_p$ is a convex risk measure while $\VaR_p$ is not. Similarly, for $X\in\X$ and $p\in(0,1]$, the \emph{left-ES} risk measure (see e.g., \cite{EWW15}) is defined by
$$\ES_p^-(X)=\frac{1}{p}\int^p_0\VaR_t(X)\d t.$$
Throughout the paper, we write $x\wedge y=\min\{x,y\}$, $x\vee y=\max\{x,y\}$, $x_+=x\vee 0$ and $x_-=(-x)\vee 0$. For an event $A\in\mathcal F$,   its complement is denoted by $A^c$.

\section{Optimal insurance contract design}
\label{sec:design}

In this section, we explain the optimal insurance design problem. 
For the economic setting, 
we make the following assumptions: 
\begin{enumerate}[(A)] 
\item The insured and the insurer may hold different attitudes towards risk. The insured adopts the risk measure $\rho:\X\to\R$ while the insurer uses the risk measure $\psi:\X\to\R$. The insured and the insurer do not observe the risk measure of their counterparty.

%

\item The premium functional is specified as 
 $\pi:\I \to\R$, which usually does not take negative values. 
For insurance loss $X\in\X_+$, note that  $ X-f(X) + \pi(f) $ is the total risk of the insured,
and $f(X) - \pi(f)$ is the total risk of the insurer. Thus, the risk values of the insurance loss to the insured and the insurer are $\rho(X-f(X)+\pi(f))$ and $\psi(f(X)-\pi(f))$, respectively.

\item The insured and the insurer agree on an insurance contract $f\in\I$ that is \emph{Pareto optimal} defined next.  
\end{enumerate}

\begin{definition}
For $X\in\X_+$, $\pi:\I\to\R$, and $\rho,\psi:\X\to\R$, an insurance contract $f\in\I$ is called Pareto optimal if there is no $g\in\I$, such that
$$\rho(X-f(X)+\pi(f))\ge\rho(X-g(X)+\pi(g))$$
and
$$\psi(f(X)-\pi(f))\ge\psi(g(X)-\pi(g)),$$
with at least one of the two inequalities strict. Pareto optimality  is also known as (Pareto)  \emph{efficiency}.
\end{definition}

A Pareto optimization problem is closely related to the minimization of a convex combination of the objective functionals of all parties, which can be seen in, e.g., \cite{G74}, \cite{BS08},   \cite{CLW17} and \cite{ELW18}. For $X\in\X_+$, $\pi:\I\to\R$, and $\rho,\psi:\X\to\R$, we define the set of minimizers of the sum of the two objectives for the insured and the insurer as
\begin{align*}
\I_{\rho,\psi}^X=\argmin_{g\in\I}\{\rho(X-g(X)+\pi(g))+\psi(g(X)-\pi(g))\}.
\end{align*}
If we further assume that $\rho$ and $\psi$ are translation invariant, then we have
\begin{align}\label{eq:set_min}
\I_{\rho,\psi}^X=\argmin_{g\in\I}\{\rho(X-g(X))+\psi(g(X))\}.
\end{align}
In this case, the set $\I_{\rho,\psi}^X$ is independent of the choice of the premium functional $\pi$.
Below we give a characterization of the Pareto-optimal problem in our context as the minimization of the total insurance value of the insured and the insurer.

\begin{proposition}\label{prop:pareto}
For two translation-invariant risk measures $\rho,\psi:\X\to\R$ and $X\in\X_+$, the following are equivalent: 
\begin{enumerate}[(i)]
\item  an insurance contract $f\in\I$ is Pareto optimal for all $\pi:\I\to\R_+$;
\item  an insurance contract $f\in\I$ is Pareto optimal for $\pi:h\mapsto \psi(h(X))$;
\item
$f\in\I_{\rho,\psi}^X$. 
\end{enumerate}
 
\end{proposition}

Proofs of all results in this paper are in Section \ref{sec:tech}.

In a similar spirit to Proposition \ref{prop:pareto}, a characterization of Pareto optimality in the context of risk sharing problems can be found in \cite{ELW18}. Proposition \ref{prop:pareto} ensures that if the objectives $\rho$ and $\psi$ for the two parties are translation invariant, then by \eqref{eq:set_min}, a Pareto-optimal insurance contract can typically be obtained by solving the following minimization problem:
\begin{align}\label{eq:sum_min}
\min_{g\in\I}\left\{\rho(X-g(X))+\psi(g(X))\right\}.
\end{align}

A minimizer of \eqref{eq:sum_min} may not be unique in many situations. Hence, the set $\I^X_{\rho,\psi}$ of efficient ceded loss functions is not a singleton in general. In the literature on optimal insurance design problems, there are many common sets of ceded loss functions.
Some notable refinements include: 
\begin{enumerate}
\item The set $\mathcal I_0$ of all non-negative functions $f$  on $[0,\infty)$ satisfying $f(x)\le x$ for $x\ge 0$. This property means that the payment cannot exceed the total loss incurred, and it is a common feature of almost all insurance contracts in practice. In particular, $f(0)=0$, and thus there is no insurance payment if there is no loss incurred. 
\item The set $\mathcal I_1$ of all increasing functions in $\mathcal I_0$. This property means that larger incurred losses  lead to higher payments to the insured. 
\item The set $\mathcal I_2=\{f\in\I_1:f(y)-f(x)\le y-x \mbox{ for all $y \ge x\ge 0$}\},$ 
which is the set of all \emph{slowly growing} increasing functions in $\mathcal I_1$. The slowly growing property is commonly assumed to avoid the problem of ex-post moral hazard (\cite{HMS83}) via the concept of comonotonicity; see Proposition \ref{lem:1} below. 

\item The set 
%
$\mathcal I^d_1=\{f\in  \I_1 : f(x) \le (x-d)_+ \mbox{ for all }x\ge 0\}$. Ceded loss functions within this set does not exceed the direct deductible form. Note that $$\I^d_1=\{f\in\I_1:f(d)=0,~x-f(x)\ge d ~\text{for all}~ x>d\}.$$
 Thus this class includes contract functions with deductible $d\ge 0$. Also, we require that the retained loss of the insured should be at least at the deductible level $d$, given that the random loss exceeds the deductible level.
In particular, we have $\I^0_1=\I_1$.

\end{enumerate}
Among the above sets, we have 
$$
\I_2\subset \I_1 \subset \I_0 \mbox{~~~and~~~} \I_1^d \subset \I_1\subset \I_0.
$$
Throughout, $\subset$ represents non-strict set inclusion. 
Contracts of deductible forms within the set $\I^d_1$ are commonly seen in the insurance market. We next give some examples.
\begin{example}[Deductible insurance with coinsurance]
Consider the following ceded loss function:
$$f(x)=\alpha(x-d)_+,~~x\ge 0,$$
which presents an insurance contract with deductible $d\ge 0$ and coinsurance parameter $\alpha\in[0,1]$. We have $f\in\I^d_1$ since $f$ is bounded from above by $(x-d)_+$. See Figure \ref{fig:example} (left-hand panel).
\end{example}
\begin{example}[Deductible insurance with policy limit]
The following ceded loss function
$$f(x)=(x-d)_+\wedge u,~~x\ge 0,$$
is also in the set $\I^d_1$. It represents an insurance contract truncated at deductible $d\ge 0$ and censored at the policy upper limit $u\ge 0$. The function is plotted in Figure \ref{fig:example} (right-hand panel).
\end{example}
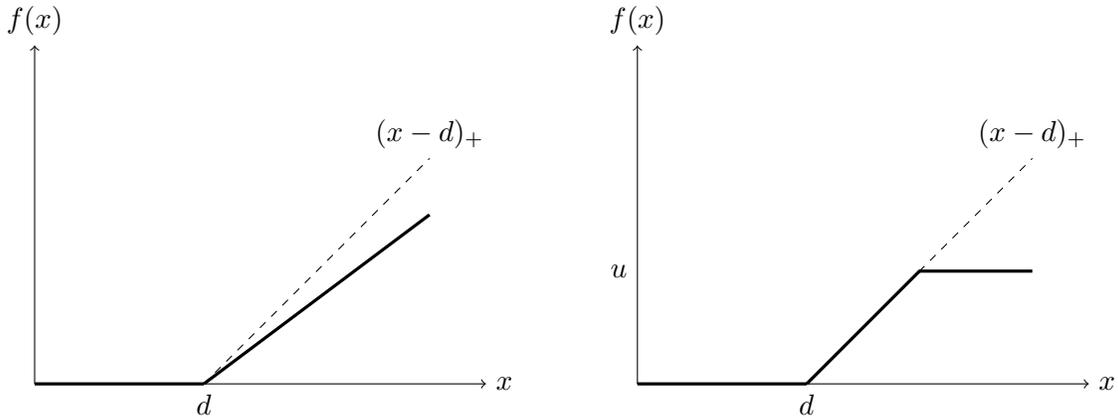
\begin{figure}[htbp]
\begin{center}

\begin{tikzpicture}[scale=1.5]
\draw[->] (0,0) -- (4,0) node[right]{$x$};
\draw[->] (0,0) -- (0,3) node[above]{$f(x)$};
\draw[very thick] (0,0) -- (1.5,0);
\draw[dashed] (0,0) -- (1.5,0) node[below]{$d$};
\draw[dashed] (1.5,0) -- (3.5,2) node[above]{$(x-d)_+$};
\draw[very thick] (1.5,0) -- (3.5,1.5);
\end{tikzpicture}~~~~~~~
\begin{tikzpicture}[scale=1.5]
\draw[->] (0,0) -- (4,0) node[right]{$x$};
\draw[->] (0,0) -- (0,3) node[above]{$f(x)$};
\draw[very thick] (0,0) -- (1.5,0);
\draw[very thick] (1.5,0) -- (2.5,1) -- (3.5,1);
\draw[dashed] (0,0) -- (1.5,0) node[below]{$d$};
\draw[dashed] (1.5,0) -- (3.5,2) node[above]{$(x-d)_+$};
\draw (0,1) node[left]{$u$};
\end{tikzpicture}

\end{center}
\caption{Solid lines represent the ceded loss functions of deductible insurance with coinsurance (left-hand panel) and deductible insurance with policy limit (right-hand panel); dashed lines represent ceded loss function with direct deductible}
\label{fig:example}
\end{figure}

Other subsets of $\mathcal I_1$, such as classes of convex functions,  piece-wise linear functions, or functions with the Vajda condition, have also been studied in the literature, but they correspond to different practical considerations. Hence, we focus on the above three subsets due to their prominence in real-world insurance contracts. 

\section{Risk measures implied by Pareto-optimal contracts}
\label{sec:char}

\subsection{Main characterization results}

In this section, we characterize measures $\rho$ and $\psi$ for the insured and the insurer in the optimal insurance design problem with different Pareto-optimal sets of ceded loss functions. 

We first collect some dependence concepts that will be helpful  to distinguish different properties of risk measures in our main results. 
A random vector $(X,Y)\in\X^2$ is said to be \emph{comonotonic} if   $(X(\omega)-X(\omega'))(Y(\omega)-Y(\omega'))\ge 0$ for almost every $\omega,\omega'\in\Omega$; see also \cite{WZ20}. A risk measure $\rho:\X\to\R$ is said to be \emph{comonotonic-additive} if $\rho(X+Y)=\rho(X)+\rho(Y)$ for all comonotonic $(X,Y)\in\X^2$. Following similar definitions as those of \cite{WZ21}, for an event $A\in\mathcal{F}$ with $0<\P(A)<1$, we call $A$ a \emph{tail event} of a random variable $X\in\X$ if $X(\omega)\ge X(\omega')$ for almost surely all $\omega\in A$ and $\omega'\in A^c$. A tail event $A$ is called a \emph{$p$-tail event} if $\P(A)=1-p$. We say that a random vector $(X_1,\dots,X_n)\in\X^n$ is \emph{$p$-concentrated} if there exists a common $p$-tail event of $X_1,\dots,X_n$. For fixed $d\ge 0$ and $p\in[0,1]$, define the sets
$$\X^d_p=\{X\in\X_+:p=\P(X\le d)\}$$
and
$$\X_p=\{X\in\X:p=\P(X\le d) \text{ for some }d\ge 0\}\supset\bigcup_{d\ge 0}\X^d_p.$$
We note that $\X_p\supset\X^d_p$ and contains random variables that can take negative values. The following proposition connects the dependence structure of  $(f(X),X-f(X))$ with
the function $f\in \I_1$.

\begin{proposition}\label{lem:1}
The following statements hold.
\begin{enumerate}[(i)]
\item $(f(X),X-f(X))$ is comonotonic for all $f\in\I_2$ and $X\in\X_+$.
\item For fixed $d>0$ and $p\in[0,1)$, $(f(X),X-f(X))$ is $p$-concentrated for all $f\in\I^d_1$ and $X\in\X^d_p$.
\end{enumerate}
\end{proposition}

Following the terminology in \cite{EMWW21}, for $\lambda\in\R$ and $p\in(0,1)$, we say that the linear combination
$$\ES^\lambda_p(X)=\lambda\ES_p(X)+(1-\lambda)\E[X], ~~X\in\X,$$
of the mean and $\ES_p$ is an \emph{$\ES/\E$-mixture}.
Note that we allow $\lambda<0$  in the definition of $\ES_p^\lambda$, so the $\ES/\E$-mixture is not necessarily a monotone risk measure. 
Define the sets $$\I_{\rho,\psi}=\bigcap_{X\in\X_+}\I^X_{\rho,\psi}\mbox{~~~and~~~}  \I^{p,d}_{\rho,\psi}=\bigcap_{X\in\X^d_p}\I^X_{\rho,\psi},$$ which are the intersections of all Pareto optimal contract sets with respect to all models of random losses in $\X_+$ and $\X^d_p$, respectively.
Different choices of $\mathcal I_{\rho,\psi}$ pin down different forms of $\rho$ and $\psi$, as we will show below. Obviously, we shall arrive at a narrower class of risk measures as the set of efficient contracts enlarges.


\begin{theorem}\label{th:1}
Suppose that $\rho$ and $\psi$ are law-invariant   convex risk measures. Then:
\begin{enumerate}[(i)]
\item $\I_{\rho,\psi}={\I_2}$ if and only if $\rho=\psi$ and $\rho$ is a convex distortion risk measure on $\X$;
\item $\I_{\rho,\psi}={\I_0}$ if and only if $\rho=\psi=\E$ on $\X$.
\end{enumerate}
\end{theorem}


Our next result, Theorem \ref{thm:ES}, establishes a relationship between deductible contracts and   ES, and it is the most sophisticated result of the present paper. The proofs of Theorems \ref{th:1} and \ref{thm:ES} are technical and   rely on additional lemmas, which are presented in Section \ref{sec:tech} together with proofs of the theorems.

\begin{theorem}\label{thm:ES}
Suppose that $\rho$ and $\psi$ are law-invariant convex risk measures with $\rho(0)=\psi(0)=0$. For any fixed $d\ge 0$ and $p\in[0,1)$, we have $\I^{p,d}_{\rho,\psi}\supset{\I^d_1}$ if and only if $\rho=\psi=\ES^\lambda_p$ on $\X_p$ for some $\lambda\ge 0$.
\end{theorem} 

We note that, given that the ceded loss functions in the set $\I_1^d$ are Pareto optimal for all insurance losses in the set $\X^d_p$, in Theorem \ref{thm:ES} we can identify the risk measure adopted by the insured and the insurer as an $\ES/\E$-mixture on a larger space of random losses $\X_p$, which does not depend on the deductible level $d$.

Theorems \ref{th:1} and \ref{thm:ES} reveal   profound  connections between common in practice sets of ceded loss functions and common classes of risk measures, as shown in Table \ref{tab:connection1}:

\begin{table}[htbp]
\centering
\begin{tabular}{ccc}
Sets of ceded loss functions & & Classes of risk measures\\\midrule
all $1$-Lipschitz ceded loss functions & $\iff$ & distortion risk measures\vspace{2mm}\\\vspace{2mm}
all non-negative ceded loss functions & $\iff$ & the mean\\
ceded loss functions with deductible form & $\iff$ & an $\ES/\E$-mixture
\end{tabular}
\caption{Connections between sets of ceded loss functions and classes of risk measures}
\label{tab:connection1}
\end{table}

As one of the most important economic interpretations of the above results, we show that
if the set of Pareto-optimal contracts between the insured and the insurer contains the set $\mathcal I_1^d$, then the risk measures of the two parties have to be an $\ES/\E$-mixture. Furthermore, if the  $\ES/\E$-mixture in Theorem \ref{thm:ES} satisfies lower semicontinuity with respect to almost sure  convergence, then it has to be an ES; see Lemma \ref{lem:mES}.

If we remove some conditions from the convex risk measures $\rho$ in Theorems \ref{th:1} and \ref{thm:ES},
then we arrive at larger classes of risk measures. 
For instance, without monotonicity in statement (i) of Theorem \ref{th:1}, we expect to arrive at the  distortion riskmetrics of \cite{WWW20a}.

\subsection{Designing insurance menus}

In this section, we discuss economic implications of our characterization results of risk measures. We assume that the risk measures $\rho$ and $\psi$ for the insured and the insurer are coherent throughout this section.

Apart from the link between the common sets of ceded loss functions and the popular classes of risk measures, it is also interesting that  all the three sets of Pareto-optimal contracts in Theorems \ref{th:1} and \ref{thm:ES} lead to the fact that the two risk measures $\rho$ and $\psi$ of the insured and the insurer are the same. In fact, when the risk measures $\rho$ and $\psi$ are coherent, a set of Pareto-optimal contracts with identical risk measures of the two parties is large enough to include all efficient contracts where the insurer is more optimistic than the insured, which can be seen from the next proposition. In this sense, the Pareto-optimal set that we obtain with identical risk measures is the union of Pareto-optimal sets with general risk measures $\rho\ge\psi$.

\begin{proposition}\label{prop:identical}
We have $\I^X_{\rho,\psi}\subset\I^X_{\psi,\psi}$ for all $X\in\X_+$ and all coherent risk measures $\rho$ and $\psi$ such that $\rho\ge\psi$.
\end{proposition}

The relation $\rho\ge\psi$ in Proposition \ref{prop:identical} indicates that the insured is more pessimistic, or more risk averse, than the insurer in the sense of \cite{P64}. Indeed, the certainty equivalent of any random loss $X$ under the preference described by the coherent risk measure $\rho$ is the risk measure $\rho(X)$ itself. Therefore, we compare risk aversion of the insured and the insurer through a direct comparison of magnitudes between coherent risk measures $\rho$ and $\psi$.

In practice, the insurer with the risk measure $\psi$ does not know the risk measure $\rho$ of the insured. Thus it is necessary for the insurer to provide a menu of contracts that is large enough to include all possible efficient contracts that might be chosen by the insured who is more pessimistic than the insurer. Specifically, we consider the following process for the design of insurance menus.

\begin{enumerate}
\item An insurer adopts the coherent risk measure $\psi$ as her own risk attitude.
\item The insurer does not have exact information about the risk attitudes of her customers. In other words, the insurer does not know the coherent risk measure $\rho$ held by any insured. However, in order to achieve the deal, the insured should be more pessimistic than the insurer (i.e.~$\rho\ge\psi$).
\item Due to incomplete information, the insurer provides a menu of contracts $\I^X_{\psi,\psi}=\bigcup_{\rho\ge\psi}\I^X_{\rho,\psi}$ for a random loss $X\in\X_+$. The set $\I^X_{\psi,\psi}$ is large enough so that Pareto optimality can be obtained for any insured that is more pessimistic than the insurer. The deal can be achieved as long as we have $\rho\ge\psi$ since both parties benefit from the final deal.
\item If the insurer aims to design a ``universal" menu of contracts so that Pareto optimality can be achieved for a bundle of random losses, the menu is then obtained by taking intersections of $\I^X_{\psi,\psi}$ with respect to a set of random losses. In this case, the insurer must choose specific classes of risk measures $\psi$, provided that the ``universal" menu of contracts contains some common sets of contracts in the insurance market. Specifically, Table \ref{tab:connection} illustrates our characterization results.
\begin{table}[htbp]
\centering
\begin{tabular}{ccc}
Pareto-optimal menu & & Insurer's risk measure $\psi$\\\midrule
$\I_{\psi,\psi}=\I_2$ & $\iff$ & $\psi$ is a distortion risk measure\vspace{2mm}\\\vspace{2mm}
$\I_{\psi,\psi}=\I_0$ & $\iff$ & $\psi=\E$\\
$\I^{p,d}_{\psi,\psi}\supset\I^d_1$ & $\iff$ & $\psi=\ES^\lambda_p$
\end{tabular}
\caption{Connections between Pareto-optimal sets of contracts and the insurer's risk measures}
\label{tab:connection}
\end{table}
\end{enumerate}

\section{Proofs  of main results and related technical lemmas}
\label{sec:tech}

In this section, we present proofs of our main results as well as several related lemmas.
As we will see, the results are technical and require highly sophisticated analysis. 
\subsection{Proofs of main results}

\begin{proof}[Proof of Proposition \ref{prop:pareto}]
\text{ }``(i)$\Rightarrow$(ii)":  This is straightforward by taking $\pi:h\mapsto \psi(h(X))$.

``(ii)$\Rightarrow$(iii)":  
Suppose that $f\in\I$ is Pareto optimal for $\pi: h\mapsto \psi(h(X))$. Assume for the sake of contradiction that $f\notin\I^X_{\rho,\psi}$. It follows that there exists $g\in\I$, such that
$$\rho(X-g(X))+\psi(g(X))<\rho(X-f(X))+\psi(f(X)).$$
By translation invariance of $\rho$ and $\psi$, we have
$$
\begin{aligned}
\rho(X-g(X)+\pi(g))&=\rho(X-g(X))+\psi(g(X))\\
&<\rho(X-f(X))+\psi(f(X))=\rho(X-f(X)+\pi(f))
\end{aligned}
$$
and
$$\psi(g(X)-\pi(g))=\psi(g(X)-\psi(g(X)))=0=\psi(f(X)-\pi(f)),$$
which leads to a contradiction to Pareto optimality of $f$. Therefore, $f\in\I^X_{\rho,\psi}$.

``(iii)$\Rightarrow$(i)": Suppose that $f\in\I^X_{\rho,\psi}$. Assume for the sake of contradiction that $f$ is not Pareto optimal for some $\pi:\I\to\R$. It follows that there exists $g\in\I$ such that
$$\rho(X-g(X)+\pi(g))\le \rho(X-f(X)+\pi(f))$$
and
$$\psi(g(X)-\pi(g))\le\psi(f(X)-\pi(f)),$$
with at least one of the above two inequalities being strict. Hence,
$$\rho(X-g(X)+\pi(g))+\psi(g(X)-\pi(g))<\rho(X-f(X)+\pi(f))+\psi(f(X)-\pi(f)),$$
which contradicts the fact that $f\in\I^X_{\rho,\psi}$. Therefore, the function $f$ is Pareto optimal for all $\pi:\I\to\R$.
\end{proof}

\begin{proof}[Proof of Proposition \ref{lem:1}]

\text{ } (i) Suppose that $f\in\I_2$. Define the function $g$ by $g(x) = x - f(x)$ for $x\in[0,\infty)$. For all $X\in\X_+$, we have $X-f(X)=g(X)$. Since $f\in\I_2$, the function $g$ is increasing and $(f(X),g(X))$ is comonotonic.

(ii) Suppose that $f\in\I^d_1$ for $d>0$. For all $X\in\X^d_p$, the set $\{X> d\}$ is a common tail event of $f(X)$ and $X-f(X)$ by the definitions of the tail event and the set $\I^d_1$. Also note that $\P(X> d)=1-p$. Therefore, $(f(X),X-f(X))$ is $p$-concentrated.
\qedhere
\end{proof}

\begin{proof}[Proof of Theorem \ref{th:1}]
	
	Let $h_0(x)= 0$ and $h_1(x)= x$, $x\ge 0$, the constant zero function and the identity, respectively.
	
	(i) ``$\Rightarrow$": Suppose that $\I_{\rho,\psi}=\I_2$. Since $h_0,h_1\in\I_2$, we have
	$$\rho(X)=\psi(X)=\min_{g\in\I}\left\{\rho(X-g(X))+\psi(g(X))\right\},~~X\in\X_+.$$
	Hence $\rho=\psi$ on $\X_+$ and $\rho(X-f(X))+\rho(f(X))=\rho(X)$ for all $f\in\I_2$ and $X\in\X_+$. By translation invariance of $\rho$ and $\psi$, we have $\rho=\psi$ on $\X$.
	
	By Proposition 4.5 of \cite{D94}, for any comonotonic $(Y,Z)\in\X_+^2$ with $Y+Z=X$, there exists $f\in \mathcal I_2$ such that $Y=f(X)$ and $Z=X-f(X)$.
	Since $X$ is arbitrary, we therefore have the equation 
	  $\rho(Y)+\rho(Z)=\rho(Y+Z)$ for all comonotonic $(Y,Z)\in\X_+^2$. Moreover, translation invariance of $\rho$ implies that $\rho(Y)+\rho(Z)=\rho(Y+Z)$ for all comonotonic $(Y,Z)\in\X^2$. This shows that $\rho$ is comonotonic-additive on $\X$.

	Moreover, we know that $\rho$ is uniformly continuous with respect to $L^\infty$-norm since $\rho$ is monetary, and $\rho$ is law invariant. Hence, $\rho$ is a convex distortion risk measure on $\X$ (see e.g., Theorem 1 of \cite{WWW20b}).
		
``$\Leftarrow$": Suppose that $\rho=\psi$ is a convex distortion risk measure on $\X$. We will prove that $\I_{\rho,\rho}=\I_2$. Since $\rho$ is a convex distortion risk measure, it is also coherent by Corollary 1 of \cite{WWW20a}. For all $f\in\I_2$ and $X\in\X_+$, we have by Proposition \ref{lem:1} that $(f(X),X-f(X))$ is comonotonic. By comonotonic-additivity of $\rho$, we have $\rho(X-f(X))+\rho(f(X))=\rho(X)$. Furthermore, due to subadditivity of $\rho$, we have $f\in\I_{\rho,\rho}$. It follows that $\I_2\subset\I_{\rho,\rho}$.

We next prove that $\I_{\rho,\rho}\subset\I_2$. For each $f\notin\I_2$, we will show that there exists $X\in\X_+$ such that $\rho(X-f(X))+\rho(f(X))>\rho(X)$. Indeed, there exists $0\le x<y$, such that $|f(y)-f(x)|>y-x$. It is clear that $f(x)\neq f(y)$. Since $\rho$ is a coherent distortion risk measure, there exists a Borel measure $\mu$ on $[0,1]$ such that $\rho=\int^1_0\ES_t\d\mu(t)$ on $\X$. Take $X=x\id_{A}+y\id_{A^c}$ where $\P(A)=1/2$. If $f(x)< f(y)$, then
$$\ES_t(X)=\left\{\begin{array}{l l}
\frac{(1-2t)x+y}{2-2t}, & 0\le t\le 1/2,\\
y, & 1/2<t<1,
\end{array}\right.$$
$$\ES_t(f(X))=\left\{\begin{array}{l l}
\frac{(1-2t)f(x)+f(y)}{2-2t}, & 0\le t\le 1/2,\\
f(y), & 1/2<t<1,
\end{array}\right.$$
$$\ES_t(X-f(X))=\left\{\begin{array}{l l}
\frac{x-f(x)+(1-2t)(y-f(y))}{2-2t}, & 0\le t\le 1/2,\\
x-f(x), & 1/2<t<1.
\end{array}\right.$$
Hence,
$$\begin{aligned}
&\frac{\rho(X-f(X))+\rho(f(X))-\rho(X)}{y-x}\\
&=\int^{1/2}_0\frac{t}{1-t}\left(\frac{f(y)-f(x)}{y-x}-1\right)\d \mu(t)+\int^{1}_{1/2}\frac{f(y)-f(x)}{y-x}-1 \d\mu(t)>0.
\end{aligned}$$
Similarly, if $f(x)>f(y)$, then we have
$$\ES_t(X)=\left\{\begin{array}{l l}
\frac{(1-2t)x+y}{2-2t}, & 0\le t\le 1/2,\\
y, & 1/2<t<1,
\end{array}\right.$$
$$\ES_t(f(X))=\left\{\begin{array}{l l}
\frac{f(x)+(1-2t)f(y)}{2-2t}, & 0\le t\le 1/2,\\
f(x), & 1/2<t<1,
\end{array}\right.$$
$$\ES_t(X-f(X))=\left\{\begin{array}{l l}
\frac{(1-2t)(x-f(x))+y-f(y)}{2-2t}, & 0\le t\le 1/2,\\
y-f(y), & 1/2<t<1,
\end{array}\right.$$
and thus
$$\begin{aligned}
&\rho(X-f(X))+\rho(f(X))-\rho(X)\\
&=\int^{1/2}_0\frac{t}{1-t}\left(f(x)-f(y)\right)\d \mu(t)+\int^{1}_{1/2}f(x)-f(y) \d\mu(t)>0.
\end{aligned}$$
Therefore, $\I_{\rho,\rho}\subset\I_2$ and thus $\I_{\rho,\rho}=\I_2$.

(ii) The ``if" part is straightforward by linearity of the mean. Hence, we prove the ``only if" part. Similar to (i), since $h_0,h_1\in\I_0$, we have by translation invariance of $\rho$ and $\psi$ that $\rho=\psi$ on $\X$. Since $\I_1\subset\I_0$ and $\X^0_0\subset\X_+$, we know from Theorem \ref{thm:ES} that $\rho(X)=\E[X]$ for all $X\in\X^0_0$. Since $X\in\X$ is bounded, we take $c>0$ such that $X+c\in\X^0_0$. It follows that $\rho(X+c)=\E[X+c]$. Hence, translation invariance of $\rho$ implies $\rho(X)=\E[X]$.
	\qedhere
\end{proof}

\begin{proof}[Proof of Theorem \ref{thm:ES}]
	``$\Leftarrow$": For all $f\in\I^d_1$, note that $(f(X),X-f(X))$ is $p$-concentrated for all $X\in\X^d_p$ by Proposition \ref{lem:1}. By $p$-additivity of $\ES_p$ (see \cite{WZ21}), we have $\ES_p(X-f(X))+\ES_p(f(X))=\ES_p(X)$ and thus $f\in\I^{d,p}_{\ES_p,\ES_p}$. Hence $\I^{d,p}_{\ES_p,\ES_p}\supset\I^d_1$.
	
	``$\Rightarrow$": It suffices to show that $\rho=\psi=\ES^\lambda_p$ on $\X^d_p$ for some $\lambda\ge 0$, and that $\rho=\psi=\ES^\lambda_p$ on $\X_p$ holds due to translation invariance of $\rho$ and $\psi$. Write $h_d(x)= (x-d)_+$, $x\ge 0$, for all $d\ge 0$ and recall that $h_0(x)= 0$, $x\ge 0$. Since $h_0,h_d\in\I^d_1$, we have
	\begin{equation}\label{eq:equal}
		\rho(X)=\rho(X\wedge d)+\psi((X-d)_+)=\min_{g\in\I}\left\{\rho(X-g(X))+\psi(g(X))\right\}
	\end{equation}
	for all $X\in\X^d_p$.
	
	We first prove the case when $d=p=0$. We know from Lemma \ref{lem:mean} that $\rho(X)=\psi(X)=\lambda\E[X]$ for some $\lambda\ge 0$ and for all $X\in\X^0_0$. Since $\rho$ is translation invariant and $X+c\in\X^0_0$ for all $X\in\X^0_0$ and $c\ge 0$, we have $$\lambda\E[X]+c=\rho(X)+c=\rho(X+c)=\lambda\E[X+c]=\lambda\E[X]+\lambda c.$$ It follows that $\lambda=1$.

	We now prove the case when $d=0$ and $p\in(0,1)$.
	We known from statement \eqref{eq:equal} that $\rho=\psi$ on $X^0_p$.
	For all $X\in\X^0_0$, we define $\phi(X)=\rho(X\id_{A})$ by taking  an event $A$ independent of $X$ with $\P(A)=1-p$ (a specific choice of $A$ does not matter since $\rho$ is law invariant). It is clear that $\phi$ is law invariant, monotone, convex and uniformly continuous with respect to $L^\infty$-norm. Note that for all $X\in\X^0_0$ and all events $B$ and $C$ independent of $X$ with $\P(B)=\P(C)=1-p$, we have $X\id_{B}\stackrel{\mathrm{d}}{=}X\id_{C}$. Hence, $\phi(X)=\rho(X\id_{B})=\rho(X\id_{C})$ and thus $\phi$ is well defined.
	Since $X\id_{A}\in\X^0_p$ and $\I^{0,p}_{\rho,\psi}\supset\I_1$, we have
	$$\begin{aligned}
		\phi(f(X))+\phi(X-f(X))&=\rho(f(X)\id_{A})+\rho((X-f(X))\id_{A})\\
		&=\rho(f(X\id_{A}))+\rho(X\id_{A}-f(X\id_{A}))=\rho(X\id_{A})=\phi(X)
	\end{aligned}$$
	for all $f\in\I_1$ and $X\in\X^0_0$. It follows from Lemma \ref{lem:mean} that $\phi(X)=\lambda\E[X]$ for some $\lambda\ge 0$ and for all $X\in\X^0_0$. For all $X\in\X^0_p$, we  take any random variable $Y$ such that $Y\stackrel{\mathrm{d}}{=}X|X>0$. We have $Y\in\X^0_0$ and $X\id_{\{X>0\}}\stackrel{\mathrm{d}}{=}Y\id_{A}$. Thus
	$$\rho(X\id_{\{X>0\}})=\rho(Y\id_{A})=\lambda\E[Y]=\lambda\ES_p(X).$$
	It follows that 
	\begin{equation}\label{eq:case1}
		\rho(X)=\psi(X)=\rho(X\id_{\{X>0\}})+\rho(X\id_{\{X=0\}})=\lambda\ES_p(X)
	\end{equation}
	for all $X\in\X^0_p$. Note that for all $X\in\X^0_p$, $$\ES_p(X)=\frac{1}{1-p}\E[X\id_{X>0}]=\frac{1}{1-p}\E[X].$$ Hence, we have
	$$\rho(X)=\lambda'\ES_p(X)+(1-\lambda')\E[X]$$
	for all $X\in\X^0_p$, where $\lambda'=(\lambda-1+p)/p$. By equation \eqref{eq:case1} and Lemma \ref{prop:char_es}, we have $\lambda\ge 1-p$ and thus $\lambda'\ge 0$.
	
	Next, we prove the case when $d>0$ and $p=0$. For all $X\in\X^0_0$, we have $X+d\in\X^d_0$. We obtain from $\I^{d,0}_{\rho,\psi}\supset\I^d_1$ that
	\begin{equation}\label{eq:zero1}
		\rho(X+d-f(X+d))+\psi(f(X+d))=\rho(X+d)
	\end{equation}
	for all $f\in\I^d_1$.
	Take those $f$ that are of the form $f(x)=g(x-d)$ for any $g\in\I_1$ and all $x\ge d$. Noting that $\rho$ is translation invariant, we have
	\begin{equation}\label{eq:zero2}
		\rho(X-g(X))+\psi(g(X))=\rho(X)
	\end{equation}
	for all $g\in\I_1$.
	Hence, $\rho(X)=\psi(X)=\lambda\E[X]=\lambda\ES_0(X)$ for some $\lambda\ge 0$ and for all $X\in\X^0_0$ by Lemma \ref{lem:mean}. Since $\rho$ is translation invariant, we have $\lambda=1$.
	
	We finally prove the case when $d>0$ and $p\in(0,1)$. For all $X\in\X^0_p$, we have $X+d\in\X^d_p$. Following similar arguments as those we used to derive equations \eqref{eq:zero1} and \eqref{eq:zero2}, we obtain
	$$\rho(X-g(X))+\psi(g(X))=\rho(X)$$
	for all $g\in\I_1$.
	Hence, $\rho(X)=\psi(X)=\lambda\ES_p(X)$ for some $\lambda\ge 1-p$ and for all $X\in\X^0_p$ by equation \eqref{eq:case1}. For all $X\in\X^d_p$, we have $(X-d)_+\in\X^0_p$. Therefore,
	$$\rho((X-d)_+)=\psi((X-d)_+)=\lambda\ES_p[(X-d)_+]=\lambda(\ES_p(X)-d).$$
Hence,
	$\rho(X\vee d)=\rho((X-d)_++d)=\lambda\ES_p(X)+(1-\lambda)d$ and $\psi(X\vee d)=\lambda\ES_p(X)+(1-\lambda)d$. By Lemma \ref{prop:char_es}, we have $\rho(X)=\psi(X)=\lambda\ES_p(X)+(1-\lambda)\ES_p^-(X)$ for all $X\in\X^d_p$. Since $$(1-p)\ES_p(X)+p\ES_p^-(X)=\E[X],$$
	we have $\rho(X)=\psi(X)=\gamma\ES_p(X)+(1-\gamma)\E[X]$, where $\gamma=1-(1-\lambda)/p\ge 0$.
\end{proof}

\begin{proof}[Proof of Proposition \ref{prop:identical}]
Take any $X\in\X_+$ and coherent risk measures $\rho,\psi:\X\to\R$ with $\rho\ge\psi$. For all $f\notin\I^X_{\psi,\psi}$, we have
$$\rho(X-f(X))+\psi(f(X))\ge\psi(X-f(X))+\psi(f(X))>\psi(X),$$
where the last inequality is due to subadditivity of $\psi$.
With $h_1(x)=x$, $x\ge 0$, which belongs to $\I$, we have $\rho(X-h_1(X))+\psi(h_1(X))=\psi(X)$ and thus
$$\min_{g\in\I}\left\{\rho(X-g(X))+\psi(g(X))\right\}\le \psi(X).$$
It follows that $f\notin\I^X_{\rho,\psi}$ and therefore $\I^X_{\rho,\psi}\subset\I^X_{\psi,\psi}$.
\end{proof}

\subsection{Technical lemmas}
In this section we have collected technical lemmas that are related to, or were needed for proving, Theorems \ref{th:1} and \ref{thm:ES}.
We note in this regard that some parts of the proofs of the main results needed characterizations without assuming translation invariance. Hence, our next lemma characterizes risk measures $\rho$ and $\psi$ without this assumption and is restricted to the space $\X^0_0$. The lemma was used in the proof of Theorem \ref{thm:ES}.
\begin{lemma}\label{lem:mean}
Suppose that risk measures $\rho$ and $\psi$ are law invariant, monotone, convex and uniformly continuous with respect to $L^\infty$-norm. Then we have the following two characterization results.
\begin{enumerate}[(i)]
\item The inclusion $$\bigcap_{X\in\X^0_0}\argmin_{g\in\I}\{\rho(X-g(X))+\psi(g(X))\}\supset\I_2$$
holds if and only if
\begin{equation}\label{eq:riskmetric}
\rho(X)=\psi(X)=\int^\infty_0h(\P(X\ge x))\d x
\end{equation}
for all $X\in\X^0_0$,
where $h:[0,1]\to[0,\infty)$ is an increasing concave function with $h(0)=0$.\footnote{Functionals of form \eqref{eq:riskmetric} belong to the family of distortion riskmetrics of \cite{WWW20a} with increasing distortion functions.}
\item The inclusion $$\bigcap_{X\in\X^0_0}\argmin_{g\in\I}\{\rho(X-g(X))+\psi(g(X))\}\supset\I_1$$
holds if and only if $\rho=\psi=\lambda\E$ on $\X^0_0$ for some $\lambda\ge 0$.
\end{enumerate}
\end{lemma}
\begin{proof}
\text{} (i) The proof is implied by that of Theorem \ref{th:1} (i).

(ii) The ``if" part is straightforward by linearity of the mean. Hence, we prove the ``only if" part. Since $\I_1\supset\I_2$, by (i), we have $\rho(X)=\psi(X)=\int^\infty_0h(\P(X\ge x))\d x$ for all $X\in\X^0_0$. By Theorem 5 of \cite{WWW20a}, there is a finite Borel measure $\mu$ on $[0,1]$ such that $\rho(X)=\int^1_0\ES_\alpha(X)\,\mu(\mathrm{d}\alpha)$ for $X\in\X^0_0$.
For all $0<\alpha\le 1$, there exists differentiable $f\in\I_1$ such that $f'(x)\le 1$ for all $x\in [0,\VaR_\alpha(X))$ and $f'(x)>1$ for all $x\in [\VaR_\alpha(X),\infty)$. Thus $x\mapsto x-f(x)$ is increasing on $[0,\VaR_\alpha(X))$ and decreasing on $[\VaR_\alpha(X),\infty)$ in strict sense. According to Lemma A.3  and Lemma A.7 of \cite{WZ21}, we have a $p$-tail event $A$ of $X$ and $f(X)$ with $$\{X>\VaR_\alpha(X)\}\subset A\subset\{X\ge\VaR_\alpha(X)\}$$ such that
	$$\ES_\alpha(X)=\E[X|A]~~\text{and}~~\ES_\alpha(f(X))=\E[f(X)|A].$$
	On the other hand, for a $p$-tail event $B$ of $X-f(X)$ satisfying $$\{X-f(X)>\VaR_\alpha(X-f(X))\}\subset B\subset\{X-f(X)\ge\VaR_\alpha(X-f(X))\},$$ we have
	$$\ES_\alpha(X-f(X))=\E[X-f(X)|B]>\E[X-f(X)|A].$$
	Thus we have
	$$\ES_\alpha(f(X))+\ES_\alpha(X-f(X))>\E[f(X)|A]+\E[X-f(X)|A]=\E[X|A]=\ES_\alpha(X)$$
	and so
	$$\begin{aligned}\rho(f(X))+\rho(X-f(X))&=\int^1_0\ES_\alpha(f(X))+\ES_\alpha(X-f(X))\,\mu(\mathrm{d}\alpha)\\&>\int^1_0\ES_\alpha(X)\,\mu(\mathrm{d}\alpha)=\rho(X),\end{aligned}$$
	which leads to a contradiction. Hence, $\mu((0,1])=0$ and $\rho(X)=\psi(X)=\lambda\E[X]$ for some $\lambda\ge 0$ and for all $X\in\X^0_0$.
\end{proof}
	
The next lemma characterizes an $\ES/\E$-mixture. The lemma 
implies that a law-invariant convex risk measure dominated by an $\ES/\E$-mixture must be the $\ES/\E$-mixture itself provided that it coincides with the $\ES/\E$-mixture somewhere.
We used the lemma when proving Theorem \ref{thm:ES}.
\begin{lemma}\label{prop:char_es}
Let $\rho:\X\to\R$ be a law-invariant convex risk measure. Fix $d\ge 0$ and $p\in(0,1)$. We have $\rho(X)=\rho((X-d)_+)+\rho(X\wedge d)$ and $\rho(X\vee d)=\lambda\ES_p(X)+(1-\lambda)d$ for all $X\in\X^d_p$ with $\lambda\in\R$ if and only if $\rho(X)=\lambda\ES_p(X)+(1-\lambda)\ES_p^-(X)$ for all $X\in\X^d_p$ with $\lambda\ge 1-p$.
\end{lemma}
\begin{proof}
The ``if" part follows immediately from the definitions of $\ES_p$ and $\ES^-_p$. Hence, we prove the ``only if" part.

Since $\rho$ is a law-invariant convex risk measure, for all $X\in\X^d_p$ we write 
$$\rho(X)=\sup_{Z\in\mathcal{Q}}\{\E[ZX]+V(Z)\},$$
where $\mathcal{Q}$ is a set of Radon-Nikodym derivatives and $V$ is a mapping from $\mathcal{Q}$ to $[-\infty,0]$
(see e.g., \cite{JST06}). We first show that $Z\le \lambda/(1-p)$ for all $Z\in\mathcal{Q}$. Assume for the sake of contradiction that $\P(Z'>\lambda/(1-p))>0$ for some $Z'\in\mathcal{Q}$. Take $A\subset\{Z'>\lambda/(1-p)\}$ and $Y=\id_A(d+1)\gamma+\id_B(d+1)$ for $\gamma>1$, where $\P(A\cup B)=1-p$ and $A\cap B=\emptyset$. It is clear that $Y\in\X^d_p$. We have
$$\begin{aligned}
\sup_{Z\in\mathcal{Q}}\{\E[ZY]+V(Z)\}&\ge\E[Z'(\id_A(d+1)\gamma+\id_B(d+1))]+V(Z')\\
&\ge (d+1)\gamma\E[Z'\id_A]+V(Z')\\&=\E[Z'|A]\E[\id_A(d+1)\gamma]+V(Z').
\end{aligned}$$
On the other hand, we have $$\begin{aligned}\lambda\ES_p(Y)+(1-\lambda)d&=\lambda\ES_p(\id_A(d+1)\gamma+\id_B(d+1))+(1-\lambda)d\\
&=\frac{\lambda}{1-p}\E[\id_A(d+1)\gamma+\id_B(d+1)]+(1-\lambda)d.\end{aligned}$$
Since $\E[Z'|A]>\lambda/(1-p)$, we have $$\lim_{\gamma\to\infty}(\E[Z'|A]\E[\id_A(d+1)\gamma]+V(Z'))>\lim_{\gamma\to\infty}(\lambda\ES_p(Y)+(1-\lambda)d),$$ which contradicts the assumption that $\rho(X)\le\lambda\ES_p(X)+(1-\lambda)d$ for all $X\in\X^d_p$. Therefore, we have $Z\le\lambda/(1-p)$ for all $Z\in\mathcal{Q}$. On the other hand, since $\E[Z]=1$, we have $\lambda/(1-p)\ge 1$ and thus $\lambda\ge 1-p$.

We next show that $\rho(X)=\lambda\ES_p(X)+(1-\lambda)\ES_p^-(X)$ for all $X\in\X^d_p$. 
Note that $\{X>d\}$ is a common $p$-tail event of $X$ and $X\vee d$. We have $\ES_p(X)=\ES_p(X\vee d)$ and $$d=\frac{1}{p}\E[(X\vee d)\id_{\{X\le d\}}]=\ES_p^-(X\vee d).$$
It follows that
$$\begin{aligned}
\sup_{Z\in\mathcal{Q}}\{\E[Z(X\vee d)]+V(Z)\}&=\rho(X\vee d)\\
&=\lambda\ES_p(X)+(1-\lambda)d=\lambda\ES_p(X\vee d)+(1-\lambda)\ES_p^-(X\vee d).
\end{aligned}$$
For $X_1,X_2,\ldots\in\X^d_p$ and $X_n\downarrow X$, since $Z$ is non-negative and bounded from above by $1/(1-p)$, the dominated convergence theorem implies $$\lim_{n\to\infty}\sup_{Z\in\mathcal{Q}}\{\E[ZX_n]+V(Z)\}=\sup_{Z\in\mathcal{Q}}\{\E[ZX]+V(Z)\},$$ which means that $\rho$ is continuous from above. Hence, $$\rho(X)=\max_{Z\in\mathcal{Q}}\{\E[ZX]+V(Z)\}$$ for all $X\in\X^d_p$; see e.g., Corollary 4.35 of \cite{FS16}. It follows that there exists $Z_0\in\mathcal{Q}$ such that \begin{equation}\label{eq:RN}
	\E[Z_0(X\vee d)]+V(Z_0)=\frac{\lambda}{1-p}\E[(X\vee d)\id_{\{X>d\}}]+\frac{1-\lambda}{p}\E[(X\vee d)\id_{\{X\le d\}}].
\end{equation}
We claim that $Z_0=\lambda\id_{\{X>d\}}/(1-p)+(1-\lambda)\id_{\{X\le d\}}/p$. Indeed, assume for the sake of contradiction that $Z_0\neq\lambda\id_{\{X>d\}}/(1-p)+(1-\lambda)\id_{\{X\le d\}}/p$. Since
$$\E[Z_0]=1=\E\left[\frac{\lambda}{1-p}\id_{\{X>d\}}+\frac{1-\lambda}{p}\id_{\{X\le d\}}\right],$$
we have $$\P\left(\left(Z_0-\frac{\lambda}{1-p}\id_{\{X>d\}}-\frac{1-\lambda}{p}\id_{\{X\le d\}}\right)_+>0\right)>0$$ and $$\P\left(\left(Z_0-\frac{\lambda}{1-p}\id_{\{X>d\}}-\frac{1-\lambda}{p}\id_{\{X\le d\}}\right)_->0\right)>0.$$
Note that $\lambda/(1-p)\ge 1\ge (1-\lambda)/p$. Hence,
$$\left\{\left(Z_0-\frac{\lambda}{1-p}\id_{\{X>d\}}-\frac{1-\lambda}{p}\id_{\{X\le d\}}\right)_+>0\right\}\subset\{X\le d\}.$$
We also note that
$$\left\{\left(Z_0-\frac{\lambda}{1-p}\id_{\{X>d\}}-\frac{1-\lambda}{p}\id_{\{X\le d\}}\right)_->0\right\}\cap\{X>d\}\neq\emptyset.$$
Otherwise, we must have $Z_0=\lambda/(1-p)$ and
$$\P\left(\left(Z_0-\frac{\lambda}{1-p}\id_{\{X>d\}}-\frac{1-\lambda}{p}\id_{\{X\le d\}}\right)_->0\right)=0,$$
which leads to contradiction.
These considerations imply that
$$\begin{aligned}&\E\left[\left(Z_0-\frac{\lambda}{1-p}\id_{\{X>d\}}-\frac{1-\lambda}{p}\id_{\{X\le d\}}\right)(X\vee d)\right]\\&=\E\left[\left(Z_0-\frac{\lambda}{1-p}\id_{\{X>d\}}-\frac{1-\lambda}{p}\id_{\{X\le d\}}\right)_+(X\vee d)\right]\\
&\quad-\E\left[\left(Z_0-\frac{\lambda}{1-p}\id_{\{X>d\}}-\frac{1-\lambda}{p}\id_{\{X\le d\}}\right)_-(X\vee d)\right]\\
&<d\left(\E\left[\left(Z_0-\frac{\lambda}{1-p}\id_{\{X>d\}}-\frac{1-\lambda}{p}\id_{\{X\le d\}}\right)_+\right]-\E\left[\left(Z_0-\frac{\lambda}{1-p}\id_{\{X>d\}}-\frac{1-\lambda}{p}\id_{\{X\le d\}}\right)_-\right]\right)\\
&=0,\end{aligned}$$
which contradicts equation \eqref{eq:RN}. Therefore, we must have $Z_0=\lambda\id_{\{X>d\}}/(1-p)+(1-\lambda)\id_{\{X\le d\}}/p$. Hence, $Z_0=\lambda\id_{\{X>d\}}/(1-p)+(1-\lambda)\id_{\{X\le d\}}/p\in\mathcal{Q}$ and $V(Z_0)=0$. It follows that $$\sup_{Z\in\mathcal{Q}}\{\E[ZX]+V(Z)\}\ge\E\left[\frac{\lambda}{1-p}X\id_{\{X>d\}}+\frac{1-\lambda}{p}X\id_{\{X\le d\}}\right]=\lambda\ES_p(X)+(1-\lambda)\ES_p^-(X).$$
On the other hand, we have $$\rho(X)\le\lambda\ES_p(X)+(1-\lambda)d=\gamma\ES_p(X)+(1-\gamma)\ES_p^-(X),$$
for some $1-p\le\lambda\le\gamma\le 1$ since $\ES^-_p(X)\le d\le \ES_p(X)$. Hence, 
there exists $\lambda\le\lambda^\prime\le \gamma$, such that $\rho(X)=\lambda^\prime\ES_p(X)+(1-\lambda^\prime)\ES_p^-(X)$.


Take $X_m=X\id_{\{X\le d\}}+(X+m)\id_{\{X>d\}}$ for $m>0$. We have $X_m\in\X^d_p$. For some $\lambda_m\in[\lambda,1]$,
\begin{equation}\label{eq:lambda1}
\rho(X_m)=\lambda_m\ES_p(X_m)+(1-\lambda_m)\ES^-_p(X_m)=\lambda_m\ES_p(X)+\lambda_m m+(1-\lambda_m)\ES^-_p(X).
\end{equation}
Since
$\rho(X_m\vee d)=\lambda\ES_p(X)+\lambda m+(1-\lambda)d,$
this implies that there exists $m>0$ such that $\lambda_m=\lambda$. Indeed, otherwise we can take $m\to\infty$ and have a contradiction to $\rho(X_m)\le \rho(X_m\vee d)$ by monotonicity of $\rho$. On the other hand, for $m$ such that $\lambda_m=\lambda$, we have 
\begin{equation}\label{eq:lambda2}
\begin{aligned}
\rho(X_m)=\rho(X_m\vee d)-d+\rho(X_m\wedge d)&=\lambda\ES_p(X)+\lambda m-\lambda d+\rho(X\wedge d)\\
&=\rho(X)+\lambda m=\lambda^\prime\ES_p(X)+(1-\lambda^\prime)\ES_p^-(X)+\lambda m.
\end{aligned}
\end{equation}
Equations \eqref{eq:lambda1} and \eqref{eq:lambda2}, together with $\lambda_m=\lambda$, yield that $\lambda^\prime=\lambda$ for all $X\in\X^d_p$. This completes the proof.
\end{proof}

Finally, we give a little lemma on   properties of $\ES/\E$-mixtures that can  precisely pin down the family of ES within the class of  $\ES/\E$-mixtures obtained in Theorem \ref{thm:ES}. 

\begin{lemma}\label{lem:mES}
For an $\ES/\E$-mixture $\rho= \lambda \ES_p+ (1-\lambda)\E$, we have the following statements:
\begin{enumerate}[(i)]
\item  $\rho$ is lower semicontinuous with respect to almost sure convergence if and only if $\lambda\ge 1$;
\item $\rho$ is convex if and only if $\lambda \ge 0$;
\item $\rho$ is monotone if and only if $\lambda \in [1-1/p,1]$. 
\end{enumerate}
In particular, $\rho$ is monotone and lower semicontinuous with respect to almost sure convergence  if and only if it is $\ES_p$.
\end{lemma}
\begin{proof} 
\text{ } (i) Suppose that $\lambda<1$. Let $X_k=-k\id_{\{U<1/k\}}$, where $U\sim \mathrm U[0,1]$.
Clearly, $X_k\to 0$ almost surely as $k\to\infty$, $\E[X_k]=-1$, and $\ES_p(X_k)=0$ for $k> 1/p$. 
Therefore, $$\liminf_{k\to\infty} \left( (1-\lambda) \E[X_k] + \lambda \ES_p(X_k)\right) =-(1-\lambda) < 0= \rho(0),$$ contradicting lower semicontinuity. 

(ii) We note that $\rho$ is a signed Choquet integral of \cite{WWW20a, WWW20b} 
with the (not necessarily increasing) distortion function
$$
h(t) = \lambda \left(\frac{t}{1-p}\wedge 1\right) + (1-\lambda) t,~~~t\in [0,1]. 
$$
By Theorem 3 of \cite{WWW20b}, $\rho$ is convex if and only if $h$ is concave. It is straightforward to verify that $h$ is concave if and only if $\lambda \ge 0$. 

(iii) By
Lemma 1 (i) of \cite{WWW20b}, $\rho$ is monotone if and only if $h$ is increasing.  Clearly, $\lambda >1$ implies that $h$ is strictly decreasing on $(1-p,1]$. 
For $\lambda\le 1$, increasing monotonicity of $h$ is equivalent to 
$$
\frac{\lambda}{1-p} + 1-\lambda \ge 0 ~~\Longleftrightarrow~~ \lambda \ge 1-\frac{1}{p}.
$$
 Hence, $\rho$ is monotone if and only if $\lambda \in [1-1/p,1]$.
 \qedhere
\end{proof}

\section{Concluding remarks}

We have considered the optimal insurance design problem in the sense of Pareto optimality. Unlike the previous studies, we have solved a characterization problem of the risk measures of the insured and the insurer given the Pareto-optimal contracts when the sum of risk measures of the insured and the insurer is minimized. We have linked the ES family,  the most popular convex risk measures, to the set of ceded loss functions with a deductible form commonly seen in insurance practice. We have not perceived ES to dominate other convex risk measures in the insurance market, since there are so many other factors that need to be taken into account. Nevertheless, given the large volume of research based on ES in insurance and actuarial science, 
 we see that the present paper brings in additional insights on why  ES is a natural risk measure to use   by  the insurer when evaluating risks in the insurance market.

We note that our characterization results can be   extended to the multi-player case with multiple insurers. This naturally links our study to the characterization of risk measures in risk sharing problems. Another potential application that can be further developed through our characterization results is that insurance companies may wish to evaluate risk attitudes of their customers based on contracts chosen from provided menus. This research direction requires more experimental studies as well as theoretical justifications. As yet another future direction, viewing the insured and the insurer as two economic agents in a competitive game, characterization problems may be explored via game theoretic approaches.

\subsection*{Acknowledgments} 
The authors would like to thank Taizhong Hu for helpful technical comments on revising the paper. RW is supported by the Natural Sciences and Engineering Research Council of Canada (RGPIN-2018-03823, RGPAS-2018-522590). RZ would like to acknowledge the financial support of the Natural Sciences and Engineering Research Council of Canada (RGPIN-2016-04452).



\begin{thebibliography}{99}

\bibitem[\protect\citeauthoryear{Arrow}{1963}]{A63}
Arrow, K. J. (1963). Uncertainty and the welfare economics of medical care. \emph{American Economic Review}, \textbf{53}(5), 941--973.

\bibitem[\protect\citeauthoryear{Artzner et al.}{Artzner et al.}{1999}]{ADEH99}
{Artzner, P., Delbaen, F., Eber, J.-M. and Heath, D.} (1999). Coherent measures of risk. \textit{Mathematical Finance}, \textbf{9}(3), 203--228.

\bibitem[\protect\citeauthoryear{Barrieu and Scandolo}{2008}]{BS08}
Barrieu, P. and Scandolo, G. (2008). General Pareto optimal allocations and applications to multi-period risks. \emph{ASTIN Bulletin}, \textbf{38}(1), 105--136.

\bibitem[BCBS(2016)]{BCBS2016}
BCBS (2016). Basel Committee on Banking Supervision (Jan.~2016).
\textit{Minimum Capital Requirements for Market Risk.}
Bank for International Settlements.
\url{https://www.bis.org/bcbs/publ/d352.pdf}

\bibitem[BCBS(2019)]{BCBS2019}
BCBS (2019). Basel Committee on Banking Supervision (Jan.~2019; revised Feb.~2019).
\textit{Minimum Capital Requirements for Market Risk.}
Bank for International Settlements.
\url{https://www.bis.org/bcbs/publ/d457.pdf}

\bibitem[\protect\citeauthoryear{Bernard and Tian}{2009}]{BT09}
Bernard, C. and Tian, W. (2009). Optimal reinsurance arrangements under tail risk measures. \emph{Journal of Risk and Insurance}, \textbf{76}(3), 709--725.


\bibitem[\protect\citeauthoryear{Braun and Muermann}{2004}]{BM04}
Braun, M. and Muermann, A. (2004). The impact of regret on the demand for insurance. \emph{Journal of Risk and Insurance}, \textbf{71}(4), 737--767.

\bibitem[\protect\citeauthoryear{Cai and Chi}{Cai and Chi}{2020}]{CC20}
Cai, J. and Chi, Y. (2020). Optimal reinsurance designs based on risk measures: A review. \emph{Statistical Theory and Related Fields}, \textbf{4}(1), 1--13.

\bibitem[\protect\citeauthoryear{Cai et al.}{Cai et al.}{2017}]{CLW17}
Cai, J., Liu, H. and Wang, R. (2017). Pareto-optimal reinsurance arrangements under general model settings. \emph{Insurance: Mathematics and Economics}, \textbf{77}, 24--37.

\bibitem[\protect\citeauthoryear{Cai and Tan}{2007}]{CT07}
Cai, J. and Tan, K. S. (2007). Optimal retention for a stop-loss reinsurance under the VaR and CTE risk measures. \emph{ASTIN Bulletin}, \textbf{37}(1), 93--112.

\bibitem[\protect\citeauthoryear{Cai et al.}{2008}]{CTWZ08}
Cai, J., Tan, K. S., Weng, C. and Zhang, Y. (2008). Optimal reinsurance under VaR and CTE risk measures. \emph{Insurance: Mathematics and Economics}, \textbf{43}(1), 185--196.




%

\bibitem[\protect\citeauthoryear{Cui et al.}{2013}]{CYW13}
Cui, W., Yang, J. and Wu, L. (2013). Optimal reinsurance minimizing the distortion risk measure under general reinsurance premium principles. \emph{Insurance: Mathematics and Economics}, \textbf{53}(1), 74--85.

\bibitem[\protect\citeauthoryear{Cummins and Mahul}{2004}]{CM04}
Cummins, J. D. and Mahul, O. (2004). The demand for insurance with an upper limit on coverage. \emph{Journal of Risk and Insurance}, \textbf{71}(2), 253--264.

\bibitem[\protect\citeauthoryear{Denneberg}{1994}]{D94}
{Denneberg, D.} (1994).
{\em Non-additive Measure and Integral.}
{Springer Science \& Business Media.}

\bibitem[\protect\citeauthoryear{Embrechts et al.}{2018}]{ELW18}
Embrechts, P., Liu, H. and Wang, R. (2018). Quantile-based risk sharing. \emph{Operations Research}, \textbf{66}(4), 936--949.

\bibitem[\protect\citeauthoryear{Embrechts et al.}{2021}]{EMWW21}
Embrechts, P., Mao, T., Wang, Q. and Wang, R. (2021). Bayes risk, elicitability, and the Expected Shortfall. \emph{Mathematical Finance}, published online. 

\bibitem[\protect\citeauthoryear{Embrechts et al.}{Embrechts et al.}{2015}]{EWW15}
{Embrechts, P., Wang, B. and Wang, R.} (2015). Aggregation-robustness and model uncertainty of regulatory risk measures.  {\em Finance and Stochastics},  \textbf{19}(4), 763--790.



\bibitem[\protect\citeauthoryear{F\"{o}llmer and Schied}{F\"{o}llmer and Schied}{2002}]{FS02} {F\"{o}llmer, H. and Schied, A.} (2002).
Convex measures of risk and trading constraints. \emph{Finance and Stochastics}, \textbf{6}(4), 429--447.



\bibitem[\protect\citeauthoryear{Follmer and Schied}{F\"ollmer and Schied}{2016}]{FS16}
F\"ollmer, H. and Schied, A. (2016). \emph{Stochastic Finance: An Introduction in Discrete Time}. Walter de Gruyter. Fourth Edition.

%
%
%

\bibitem[\protect\citeauthoryear{Gerber}{1974}]{G74}
{Gerber, H. U.} (1974). On additive premium calculation
principles. \emph{ASTIN Bulletin}, \textbf{7}(3), 215--222.


\bibitem[\protect\citeauthoryear{Gollier}{1996}]{G96}
Gollier, C. (1996). Optimum insurance of approximate losses. \emph{Journal of Risk and Insurance}, \textbf{63}(3), 369--380.



\bibitem[\protect\citeauthoryear{Gollier and Schlesinger}{1996}]{GS96}
Gollier, C. and Schlesinger, H. (1996). Arrow's theorem on the optimality of deductibles: A stochastic dominance approach. \emph{Economic Theory}, \textbf{7}(2), 359--363.


\bibitem[\protect\citeauthoryear{Han et al.}{Han et al.}{2021}]{HWWW21}   Han, X., Wang, B.,  Wang, R. and Wu, Q. (2021). Risk concentration and the mean-Expected Shortfall criterion.
\emph{arXiv}: 2108.05066.

\bibitem[\protect\citeauthoryear{Hofmann et al.}{Hofmann et al.}{2019}]{HHN19}
Hofmann, A., H\"afen, O. V. and Nell, M. (2019). Optimal insurance policy indemnity schedules with policyholders' limited liability and background risk. \emph{Journal of Risk and Insurance}, \textbf{86}(4), 973--988.

\bibitem[\protect\citeauthoryear{Huberman et al.}{Huberman et al.}{1983}]{HMS83}
Huberman, G., Mayers, D. and Smith Jr, C. W. (1983). Optimal insurance policy indemnity schedules. \emph{The Bell Journal of Economics}, 415--426.

\bibitem[\protect\citeauthoryear{Jouini et al.}{2006}]{JST06}
Jouini, E., Schachermayer, W. and Touzi, N. (2006). Law invariant risk measures have the Fatou property. \emph{Advances in Mathematical Economics}, \textbf{9}, 49--71.


\bibitem[\protect\citeauthoryear{Lo et al.}{2021}]{LTT21}
Lo, A., Tang, Q. and Tang, Z. (2021). Universally marketable insurance under multivariate mixtures. \emph{ASTIN Bulletin}, \textbf{51}(1), 221--243.




\bibitem[\protect\citeauthoryear{Pratt}{Pratt}{1964}]{P64}
Pratt, J. W. (1964). Risk aversion in the small and in the large. \emph{Econometrica}, \textbf{32}, 122--136.

\bibitem[\protect\citeauthoryear{Schlesinger}{1981}]{S81}
Schlesinger, H. (1981). The optimal level of deductibility in insurance contracts. \emph{Journal of Risk and Insurance}, \textbf{48}(3), 465--481.

\bibitem[\protect\citeauthoryear{Schlesinger}{1997}]{S97}
Schlesinger, H. (1997). Insurance demand without the expected-utility paradigm. \emph{Journal of Risk and Insurance}, \textbf{64}(1), 19--39.



\bibitem[\protect\citeauthoryear{Wang et al.}{Wang et al.}{2020a}]{WWW20a}
Wang, Q., Wang, R. and Wei, Y.  (2020a). Distortion riskmetrics on general spaces. \emph{ASTIN Bulletin}, \textbf{50}(4), 827--851.

\bibitem[\protect\citeauthoryear{Wang, Wei and Willmot}{Wang et al.}{2020b}]{WWW20b}
{Wang, R., Wei, Y. and Willmot, G.} (2020b).
Characterization, robustness and aggregation of signed Choquet integrals.
\emph{Mathematics of Operations Research}, \textbf{45}(3), 993--1015.

\bibitem[\protect\citeauthoryear{Wang and Zitikis}{2020}]{WZ20}
Wang, R. and Zitikis, R. (2020). Weak comonotonicity. \emph{European Journal of Operational Research}, \textbf{282}(1), 386--397.


\bibitem[\protect\citeauthoryear{Wang and Zitikis}{2021}]{WZ21}
{Wang, R. and Zitikis, R.} (2021). An axiomatic foundation for the Expected Shortfall. \emph{Management Science}, \textbf{67}(3), 1413--1429.

\bibitem[\protect\citeauthoryear{Wang et al.}{Wang et al.}{1997}]{WYP97}
{Wang, S., Young, V. R. and Panjer, H. H.} (1997). Axiomatic characterization of insurance prices. \emph{Insurance: Mathematics and Economics}, \textbf{21}(2), 173--183.

\bibitem[\protect\citeauthoryear{Yaari}{Yaari}{1987}]{Y87}
{Yaari, M. E.} (1987). The dual theory of choice under risk. \emph{Econometrica}, \textbf{55}(1), 95--115.



\end{thebibliography}
\end{document}